\documentclass[twocolumn,aps,prb,showpacs,superscriptaddress,longbibliography]{revtex4-1}
\usepackage{color}
\usepackage{mathtools}
\usepackage{graphicx}
\usepackage{SIunits}

\newcommand{\dograph}[3]{
	\begin{figure}
	\includegraphics[#2]{#1}
	\caption{\label{fig:#1}#3}
	\end{figure}
}

\begin{document}

\title{{Suppressed crosstalk between two-junction superconducting qubits with mode-selective exchange coupling}}

\author{A.~D.~K.~Finck}
\affiliation{IBM Quantum, IBM T.J.~Watson Research Center, Yorktown Heights, NY 10598, USA}

\author{S.~Carnevale}
\affiliation{IBM Quantum, IBM T.J.~Watson Research Center, Yorktown Heights, NY 10598, USA}

\author{D.~Klaus}
\affiliation{IBM Quantum, IBM T.J.~Watson Research Center, Yorktown Heights, NY 10598, USA}

\author{C.~Scerbo}
\affiliation{IBM Quantum, IBM T.J.~Watson Research Center, Yorktown Heights, NY 10598, USA}

\author{J.~Blair}
\affiliation{IBM Quantum, IBM T.J.~Watson Research Center, Yorktown Heights, NY 10598, USA}

\author{T.~G.~McConkey}
\affiliation{IBM Quantum, IBM T.J.~Watson Research Center, Yorktown Heights, NY 10598, USA}

\author{C.~Kurter}
\affiliation{IBM Quantum, IBM T.J.~Watson Research Center, Yorktown Heights, NY 10598, USA}

\author{A.~Carniol}
\affiliation{IBM Quantum, IBM T.J.~Watson Research Center, Yorktown Heights, NY 10598, USA}

\author{G.~Keefe}
\affiliation{IBM Quantum, IBM T.J.~Watson Research Center, Yorktown Heights, NY 10598, USA}

\author{M.~Kumph}
\affiliation{IBM Quantum, IBM T.J.~Watson Research Center, Yorktown Heights, NY 10598, USA}

\author{O.~E.~Dial}
\affiliation{IBM Quantum, IBM T.J.~Watson Research Center, Yorktown Heights, NY 10598, USA}

\date{\today}

\begin{abstract}

Fixed-frequency qubits can suffer from always-on interactions that inhibit independent control.  Here, we address this issue by experimentally demonstrating a superconducting architecture using qubits that comprise of two capacitively-shunted Josephson junctions connected in series.  Historically known as tunable coupling qubits (TCQs), such two-junction qubits support two modes with distinct frequencies and spatial symmetries.  By selectively coupling only one type of mode and using the other as our computational basis, we greatly suppress crosstalk between the data modes while permitting all-microwave two-qubit gates.

\end{abstract}

\maketitle


\section{Introduction}

It is an open question how to create a scalable quantum computer.  While superconducting qubits offer a promising path, cross-talk between such qubits that inhibit independent control or generate unwanted entanglement is a concern for these systems \cite{PRXQuantum.1.020318}.  One strategy is to employ flux-tunable qubits \cite{Nature.460.240, PhysRevLett.125.120504} in order to control the detuning and thus interaction strength between adjacent qubits;  however, this approach is susceptible to flux noise, which limits qubit coherence time.  Alternatively, one could employ flux-tunable couplers \cite{PhysRevApplied.10.054062, PhysRevApplied.12.054023, Stehlik2021} to dynamically alter the interaction between qubits.  One could also use a hybrid system of two types of qubits to suppress static ZZ interactions, which can impair the fidelity of simultaneous single-qubit gates by causing the transition frequency of one qubit to be dependent on the state of an adjacent qubit.  Finally, a coupler has been recently demonstrated \cite{Kandala2020} to permit all-microwave entangling gates between qubits while significantly reducing static ZZ interactions.  However, most of these strategies with fixed-frequency transmons require a high degree of frequency control during fabrication in order to produce qubits within a regime where both fast two-qubit gates are possible and unwanted cross-talk is suppressed.  There are various efforts to achieve this precision, such as laser-tuning of Josephson junctions immediately prior to cryogenic measurement \cite{Hertzberg2020, Zhang2021}.

Here, we experimentally demonstrate cross-talk cancellation by making use of two-junction qubits \cite{PhysRevLett.103.150503}.  Sometimes referred to as tunable coupling qubits (TCQs) \cite{PhysRevLett.106.030502, PhysRevB.84.184515, QI.3.1},  they consist of two capacitively-shunted Josephson junctions connected in series.  They possess two modes of excitations with distinct frequencies and spatial symmetries: a low-frequency ``bright" mode (with non-zero dipole moment) and a high-frequency ``dark" mode (lacking a dipole moment).  Here, we will refer to the ``bright" mode as the A mode and the ``dark" mode as the B mode.  Because of their different spatial symmetries, it is possible to create exchange coupling between a bus resonator and only one mode  \cite{PhysRevLett.106.030502, PhysRevLett.106.083601}.   For example, capacitively coupling to the middle pad of a TCQ will generate exchange coupling only to the B mode, assuming that the two Josephson junctions have equal critical currents and equal shunting capacitances.  We propose to store qubit information in one mode (e.g. A mode) and couple bus resonators to the other mode (e.g. B mode).   Here, unless stated otherwise, we will denote the ground state of a TCQ as $|0\rangle$ and the first excited state of the A mode as $|1\rangle$.   The lack of exchange interactions between A modes is found to suppress cross-talk, such as static ZZ interactions.  To implement two-qubit gates, we take advantage of the fact that a sizable longitudinal interaction can exist between the bus resonators and the data modes without direct exchange coupling \cite{PhysRevLett.106.030502, PhysRevApplied.7.054025}.  This allows us to realize a resonator induced phase (RIP) gate by driving the bus resonator \cite{PhysRevA.91.032325, PhysRevLett.117.250502}.

\section{Circuit Diagram and Numerical Simulations}

We illustrate our sample geometry as a circuit diagram in Fig.~\ref{fig:figure1}a.  Here, Q1 and Q2 represent the two TCQs with a bus resonator coupling to each of their middle capacitor pads.  If each of the Josephson junctions for a given TCQ have equal critical currents and are shunted by equal capacitances (denoted as $C_1$ and $C_2$ in  Fig.~\ref{fig:figure1}a), then the bus resonator will have nonzero exchange coupling with only the high frequency B mode of the TCQ.  Breaking the symmetry between the two junctions for a given TCQ will lead to some exchange coupling between the resonator and the A mode; however, if the asymmetry is not too large then cross-talk between A modes of two neighboring TCQs is negligible.

To demonstrate the robustness of cross-talk cancellation in our geometry, we performed numerical calculations of static ZZ interactions between the A modes of the two TCQs.  We quantized the circuit using standard methods \cite{Vool2017, Kjaergaard2020, Krantz2020}, treating each component in a truncated charge basis \cite{BishopThesis}.  For each TCQ, we use a Hamiltonian of the form:
\begin{equation}
\begin{split}
H_{TCQ} =& \frac{4 \pi e^2}{h} \left[g_{11} (\hat{n}_1 - n_{g1}) ^ 2 + g_{22}  (\hat{n}_2 - n_{g2}) ^ 2\right]\\
& -  \frac{8 \pi e^2}{h} g_{12} \hat{n}_1 \hat{n}_2\\
&- E_{J1} \cos{\hat{\phi}_1} - E_{J2} \cos{\hat{\phi}_2},
\end{split}
\end{equation}
where $\hat{n}_{1,2}$, $n_{g1,2}$, and  $E_{J1,2}$ are (respectively) the charge operators, charge offsets, and Josephson energies for the individual junctions.  Here, $g_{ij}$ are elements of the inverted capacitance matrix for the entire circuit \cite{Vool2017}.  We note that we perform a change in variables to transform from absolute charges in each circuit node to relative charge differences across each junction.  For the uncoupled case, we find that:
\begin{equation}
\begin{split}
g_{11} &= \frac{C_S + C_2}{2 \left[ C_1 C_2 + C_S (C_1 + C_2)  \right]},\\
g_{22} &= \frac{C_S + C_1}{2 \left[ C_1 C_2 + C_S (C_1 + C_2)  \right]},\\
g_{12} &= \frac{C_S }{ \left[ C_1 C_2 + C_S (C_1 + C_2)  \right]},
\end{split}
\end{equation}
where $C_S$ is the capacitance shunting the top and bottom capacitor pads of the TCQ.  We quantize the individual TCQs by using number operators truncated to a maximum value of $\pm 8$, diagonalize the TCQ Hamiltonian, and then truncate the resulting TCQ Hamiltonian and number operators to a subspace of the 8 lowest energy levels.  Expanding further yields negligible changes in energy levels.  We can then diagonalize the Hamiltonian for the coupled system,
\begin{equation}
\begin{split}
H_{total} &= H^{'}_{TCQ,1} + H^{'}_{TCQ,2} + H_{bus} \\
               &+ H_{int,1} + H_{int,2},\label{eqn:Htotal}
\end{split}
\end{equation}
where $H^{'}_{TCQ,i}$ is the truncated Hamiltonian for TCQ $i$, $H_{bus}$ is the Hamiltonian for the bus resonator (modeled as an ideal harmonic oscillator), and the interaction terms are given by:
\begin{equation}
H_{int,i} =  \frac{8\pi e^2}{h} (g_{int,i}) (\hat{n}_{bus}) (\hat{n}^{'}_{1,i} - \hat{n}^{'}_{2,i}),
\end{equation}
where $\hat{n}_{bus} $ is the charge operator for the bus resonator, $\hat{n}^{'}_{1,i} , \hat{n}^{'}_{2,i} $ are the truncated charge operators for TCQ $i$, and $g_{int,i}$ is the interaction term between the bus and TCQ $i$ derived from inverting the capacitance matrix of the circuit.  Equation \ref{eqn:Htotal} can be diagonalized and the static ZZ interactions between the A modes of the TCQs are calculated according to:
\begin{equation}
ZZ = \frac{1}{2} ( E_{11} + E_{00} - E_{10} - E_{01}),
\end{equation}
where $E_{ij}$ is the energy of state in the coupled system that has the greatest overlap with a tensor product of uncoupled states corresponding to TCQ 1 being in its $i$th state, TCQ 2 being in its $j$th state, and the bus being in its ground state.  For $i=1$, this corresponds to the first excited state of the A mode of a TCQ.

Next, we calculated static ZZ between the A modes of two TCQs for a range of detuning between the A mode frequencies, using the shunting capacitances of $C_1 = C_2 = 45$ fF and $C_S = 20$ fF.  We kept the junction critical currents for one TCQ fixed at values that corresponded to a longitudinal coupling of $\chi/2 = 1$ MHz between its A mode and the bus resonator, which is tuned to have a resonant frequency of 6 GHz.  We then varied the critical currents of the other TCQ and calculated the static ZZ coupling.  We deliberately introduced asymmetry between the critical currents of junction 1 and junction 2 for each TCQ in order to model realistic variations in junction size \cite{Hertzberg2020, Zhang2021}.  We plot the resulting ZZ versus detuning curves for multiple values of asymmetry in Fig.~\ref{fig:figure1}b.  We also plot ZZ versus detuning for a conventional transmon - resonator - transmon system, in which the first transmon has its frequency tuned so as to also have a longitudinal coupling of $\chi/2 = 1$ MHz with the bus and the transmons both have shunting capacitances of 70 fF.  We find that the static ZZ between the TCQ A modes is orders of magnitude lower than the transmon - transmon ZZ for a wide range of detuning, even when there is a 10\% mismatch between the Josephson junction critical currents for a given TCQ.  At high detuning (beyond 0.5 GHz), the ZZ coupling will continue to rise as the A mode of one TCQ becomes resonant with the 1 to 2 transition of the B mode of the other TCQ or with the bus resonator.  We still expect the static ZZ coupling between the A modes of the TCQs to be significantly lower than the case of two transmons in this regime.

\dograph{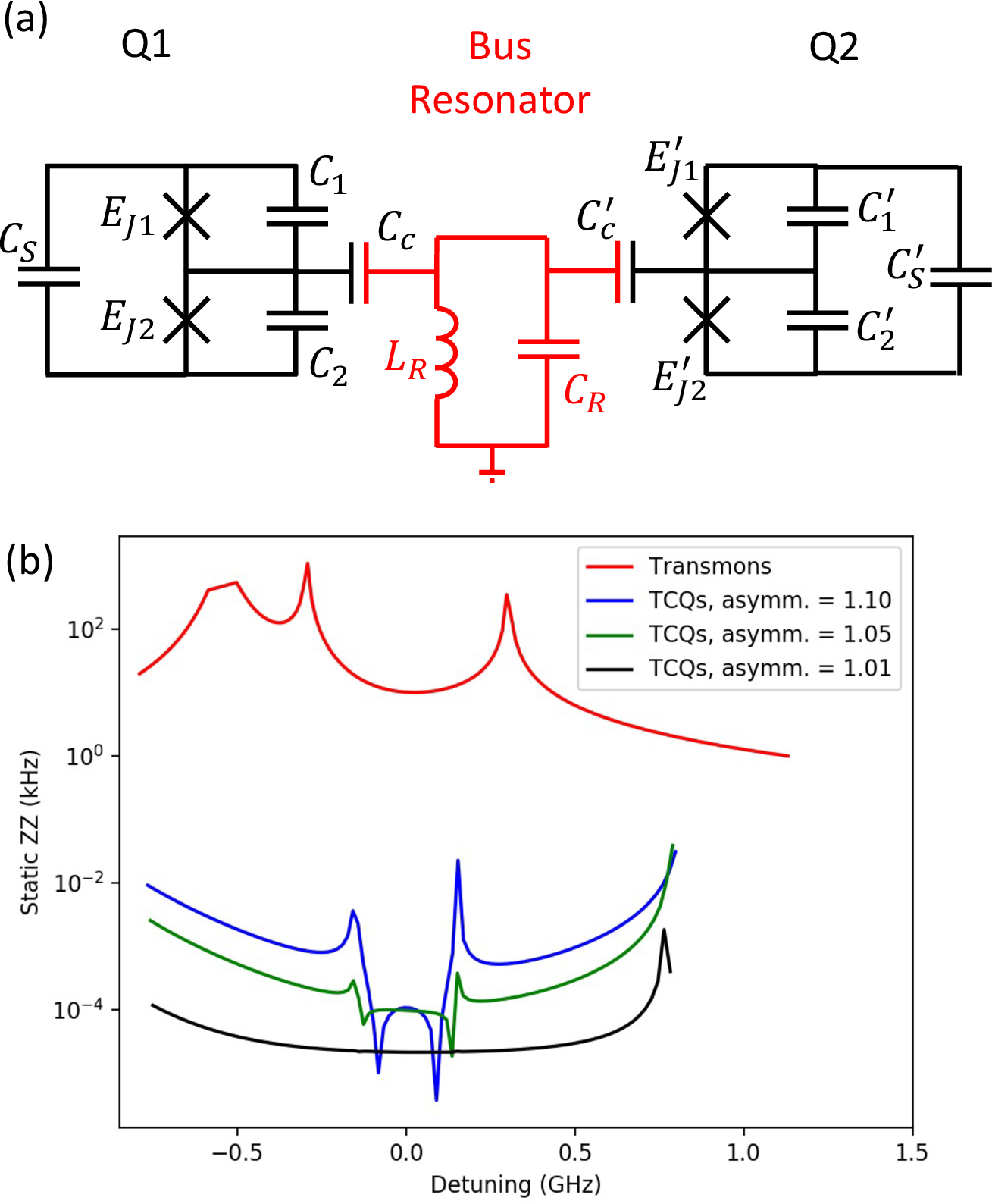}{width = 3.25 in}{(a) Circuit diagram of geometry comprising of two TCQs coupled to a bus resonator.  (b)  Numerical simulation of static ZZ interaction between data modes of either coupled transmon systems or coupled TCQ systems.  In all cases, the frequency of one qubit is kept constant while the other is varied.  For the coupled TCQ systems, we include a variable amount of asymmetry in the critical currents of the two Josephson junctions of both TCQs.}

\section{Experimental Methods}

Motivated by the robust suppression in static ZZ, we fabricated and studied a number of devices with 6 TCQs connected by bus resonators, following the circuit in Fig.~\ref{fig:figure1}a and shunting capacitances similar to the ones in the simulated TCQs of Fig.~\ref{fig:figure1}b.  We fabricated our devices on high-resitivity Si substrates.  The large features of our superconducting circuits are formed by sputtering 200 nm of Nb on the substrates, patterning with optical lithography, and dry etching with RIE.  Our Al/AlO$_x$/Al Josephson junctions are patterned with e-beam lithography, followed by a standard Dolan bridge technique \cite{Dolan1977} with angled evaporation.  Prior to aluminum evaporation, ion milling is used to remove the oxide layer of the Nb and ensure good electrical contact.

Our samples were cooled in a commercial dilution refrigerator with a base temperature of 20 mK.  We employed heavily attenuated input lines.  For the output lines, we used commercial isolators at the mixing chamber plate, followed by superconducting Nb coaxes that lead to commercial HEMTs at the 4K plate.  At room temperature, we used custom IBM electronics for waveform generation, mixing, demodulation, and digitization.

\section{Sample Properties}

A diagram for one of the devices is shown in Fig.~\ref{fig:figure2}; all experimental data are derived from this device, although qualitatively similar results were seen in others.  Each box represents a TCQ; the two frequencies in each box give the frequencies for the A mode (smaller frequency) and B mode (larger frequency) of each TCQ.  The connecting black lines between the boxes represent bus resonators that are selectively coupled to the B modes of the TCQs.  We employed coplanar waveguide bus resonators with frequencies between 5.9 and 6.3 GHz.  Between each qubit pair we print the measured static ZZ interaction between the A modes of the corresponding qubits.  We observe that the ZZ values are consistently low, being statistically indistinguishable from zero.  In terms of single-qubit performance, we found that the A modes typically had $T_1 \approx 82$ $\mu$s, $T^{echo}_2 \approx 22$ $\mu$s, and single-qubit gate errors of $\approx 3.7 \times 10^{-4}$.  We measure the same single-qubit gate errors whether the gates are run independently or simultaneously on all of the qubits.  For additional details of this device (including qubit frequencies and coherence times), see Tables \ref{table:deviceA}, \ref{table:deviceB}, and \ref{table:buses}.

\dograph{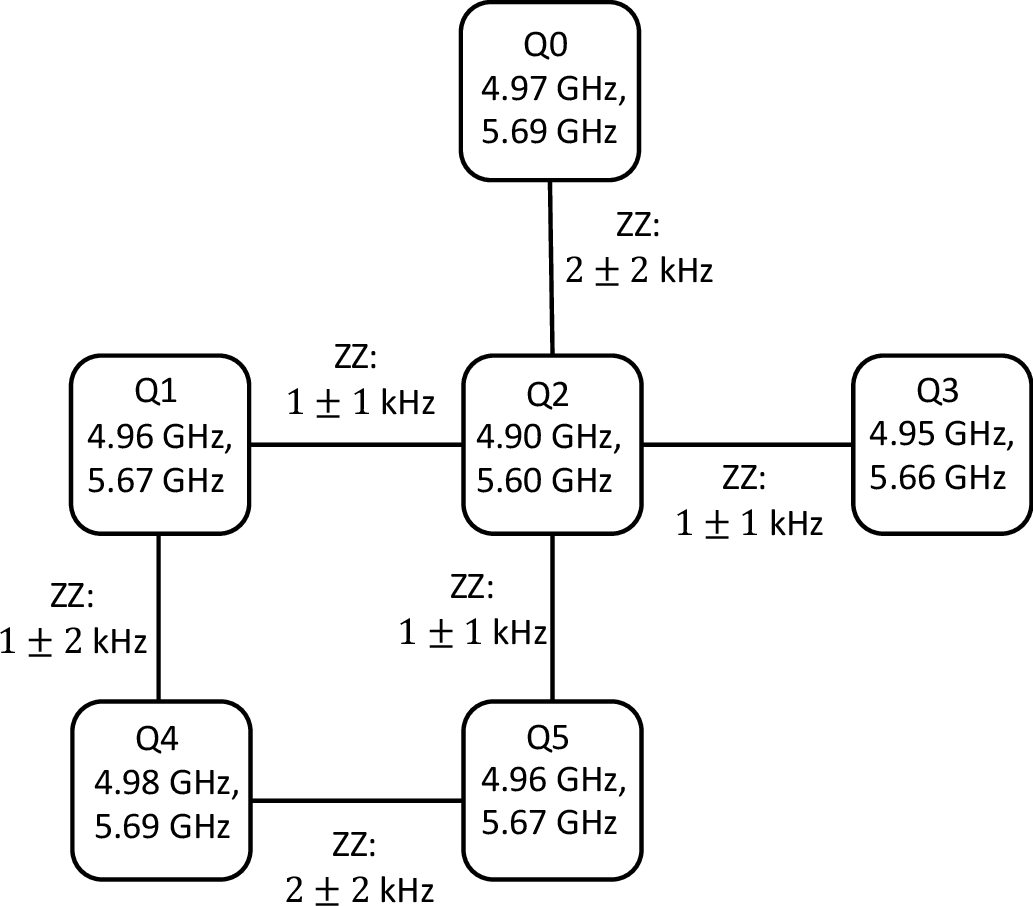}{width = 3.5 in}{Diagram of featured device.  The two frequencies in each box denote the A (lower) and B (higher) mode frequencies.  Bus resonators are denoted by black lines connecting the qubits, with the static ZZ interaction between the A modes of the connected qubits printed near the buses.}

\begin{table}[htp]
\caption{A Mode parameters.  $f_{01}$ is the transition frequency, $\alpha$ is the anharmonicity, EPG is the error per single qubit gate, $f_r$ is the frequency of the readout resonator, and $\chi$ is the longitudinal coupling between the A mode and the readout resonator.}
\begin{center}
\begin{tabular}{|c|cccccccc|}
\hline
Qubit & $f_{01}$ & $\alpha$  & $T_1$      & $T_2$      & EPG & $f_r$ & $\chi$ & Readout \\
         & (GHz)      &  (MHz)     & ($\mu$s) &  ($\mu$s) &  ($\times 10^{-4}$) & (GHz) &  (kHz) & fidelity\\
\hline
0        &  4.972     & -204          & 82         & 55           & 2.0 &  7.182 & 384& 58\%\\
1        &  4.954     & -196        & 54          & 21            & 5.2 &   7.081  &533 & 87\%\\
2        &  4.903    & -199         & 109           & 24            & 1.9 &  7.264 & 200 & 58\%\\
3        &  4.948     & -200         & 95           & 46            & 2.3 &   7.078  & 571  & 64\%\\
4        &  4.979     & -195        & 69          & 21            & 5.5 &   7.277  &865  & 84\%\\
5        &  4.959     & -195          & 81           & 12          & 6.3 &   7.169  & 528 & 91\%\\
\hline
\end{tabular}
\end{center}
\label{table:deviceA}
\end{table}%

\begin{table}[htp]
\caption{B Mode parameters.  $f_{01}$ is the transition frequency, $\alpha$ is the anharmonicity, and EPG is the error per single qubit gate.  ``n/a" denoted unmeasurable.}
\begin{center}
\begin{tabular}{|c|ccccc|}
\hline
Qubit & $f_{01}$ & $\alpha$  & $T_1$      & $T_2$      & EPG\\
         & (GHz)      &  (MHz)     & ($\mu$s) &  ($\mu$s) &  ($\times 10^{-4}$)\\
\hline
0        &  5.692     & -193         &  40        & 39          & 8 \\
1        &  5.669     & -188        & 38          & 19            & 15 \\
2        &  5.602     & -157        & 100           & n/a            & 12 \\
3        &  5.661     & n/a        & 75          & 67           & 10 \\
4        &  5.688     & -190         & 55           & 6            & 33 \\
5        &   5.670    & -189         & 54           & 11          & 18 \\
\hline
\end{tabular}
\end{center}
\label{table:deviceB}
\end{table}%

\begin{table}[htp]
\caption{Interqubit bus parameters, with bus's resonant frequency $f_{bus}$ and the longitudinal coupling $2\chi$ with the $A$ modes of the first and second qubit.  ``n/a" denoted unmeasurable.}
\begin{center}
\begin{tabular}{|c|ccc|}
\hline
Qubit & $f_{bus}$ & First qubit's $2\chi$   &  Second qubit's $2\chi$ \\
  pair       & (GHz)      &  (MHz)     &   (MHz)\\
\hline
Q0-Q2        &  5.924     & Q0: 2.5        &    Q2: 1.5 \\
Q1-Q2        &  6.288     & Q1: 0.75        &    Q2: 0.5 \\
Q1-Q4        &  6.138     & Q1: 1.1        &    Q4: 1.0 \\
Q2-Q3        &  5.994     & Q2: 1.5        &    Q3: 1.8\\
Q2-Q5        &  6.288     & Q2: 0.8        &    Q5: 0.6\\
Q4-Q5        &  6.214     & Q4: n/a        &    Q5: 0.6 \\
\hline
\end{tabular}
\end{center}
\label{table:buses}
\end{table}%

\section{Qubit-Qubit and Qubit-Resonator Interactions}

In Fig.~\ref{fig:figure3}, we explore qubit-qubit interactions and qubit-resonator interactions.  Fig.~\ref{fig:figure3}a shows an example of a Ramsey experiment to measure the static ZZ interactions between the A modes of Q1 and Q2.  In Fig.~\ref{fig:figure3}b, we plot a measurement of the ZZ interactions between the B modes of the same two qubits.  The B modes clearly have a larger static ZZ than the A modes, thus demonstrating mode-selective coupling.

To measure the longitudinal coupling $\chi$ between the bus resonator and the TCQ modes, we performed photon number splitting spectroscopy  \cite{PhysRevA.74.042318, Nature.445.515}.  This involves doing a Rabi spectroscopy measurement immediately after a detuned microwave pulse has been applied to the bus resonator.  The detuned pulse fills the bus with photons, causing a shift in the qubit frequency equal to $2\chi n$, where  $n$ is the number of photons in the bus.  The varying number of photons results in a number of evenly spaced peaks in the Rabi spectrum, allowing for a measurement of $\chi$.  In Figs.~\ref{fig:figure3}c and d, we show the spectroscopy measurement of the A and B modes of Q2 after driving the 5.92 GHz bus connecting Q0 and Q2.  We find that the A mode has a longitudinal coupling of $2\chi = 1.5$ MHz to the bus while the B mode has $2\chi = 2$ MHz.  Thus, both modes have a sizable longitudinal coupling with the bus resonator.

\dograph{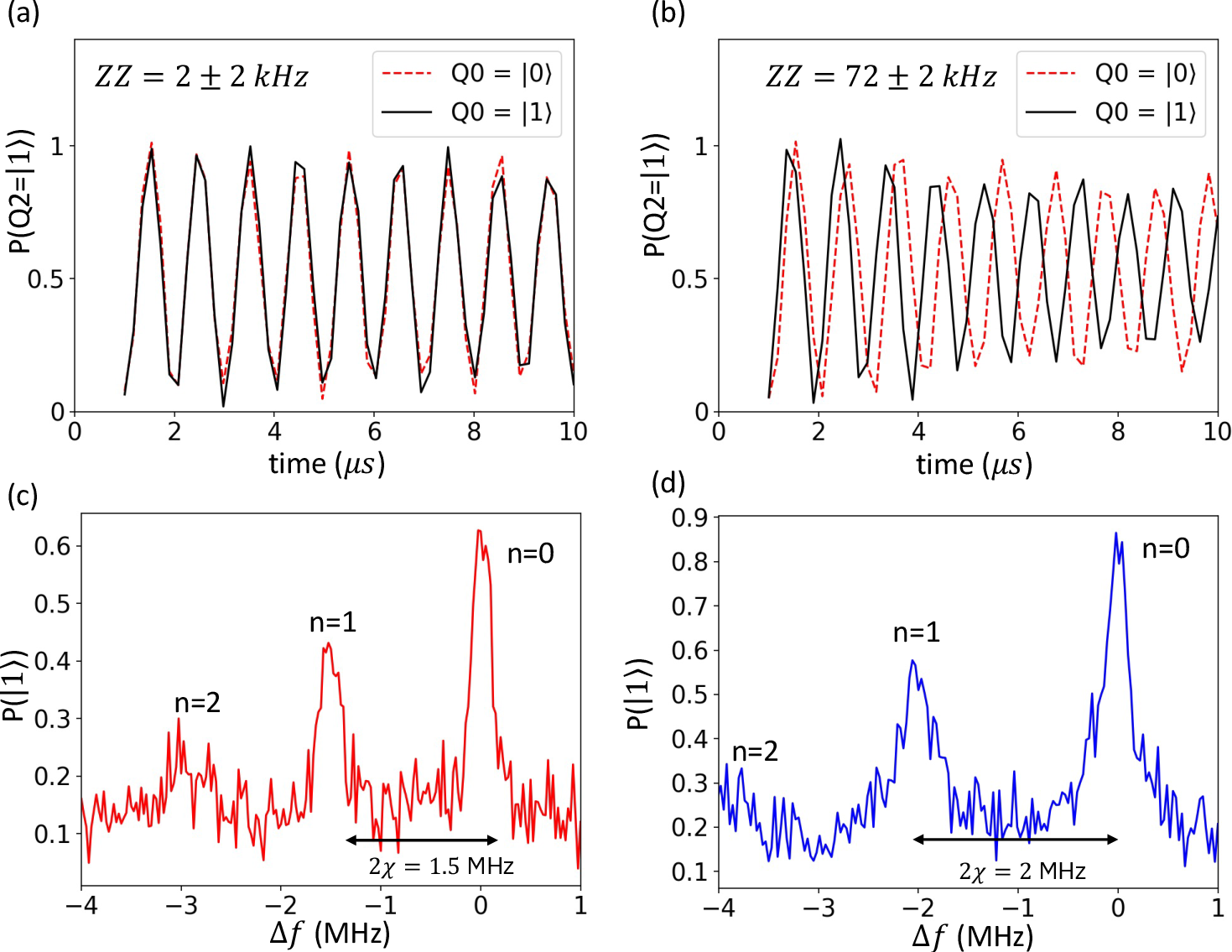}{width = 3.5 in}{(a) Measurement of static ZZ interaction between the A modes of Q0 and Q2.  (b) Measurement of static ZZ interactions between the B modes of Q0 and Q2. (c) Photon number splitting spectroscopy of the A mode of Q2 while driving the bus between Q0 and Q2.  (d) Photon number splitting spectroscopy of the B mode of Q2 while driving the bus between Q0 and Q2.  For (a) and (c), the y-axis corresponds to the probability of Q2 being the first excited state of its A mode.  For (b) and (d), the y-axis corresponds to the probability of Q2 being the first excited state of its B mode.  }

\section{RIP Gate}

Finally, we demonstrated an entangling gate between the A modes of the TCQ by means of the echoed RIP gate \cite{PhysRevA.91.032325, PhysRevLett.117.250502}.  This gate consists of a detuned pulse on the bus resonator of duration $\tau$, followed up a pair of echo pulses on A modes of the two TCQs, and then followed by another detuned bus pulse of duration $\tau$.  A Ramsey experiment can reveal the Z rotations induced by the gate, which result from state-dependent Stark shifts of the two qubits.  In Fig.~\ref{fig:figure4}a, we show examples of Z rotations of the A mode of Q2 when driving the bus resonator between Q0 and Q2, as a function of the individual bus pulse duration (i.e. $\tau$).   The black trace in Fig.~\ref{fig:figure4}a corresponds to a bus drive that is detuned by $-20$ MHz from the bus's resonant frequency while the red trace corresponds to applying no drive to the bus.  We observed that while no Z rotations occur in the absence of a bus drive, clear Ramsey oscillations are observed in the presence of the bus drive.  We note that all of the y-axes in Fig.~\ref{fig:figure4} correspond to the probability of finding a qubit in the first excited state of its A mode.

In Fig.~\ref{fig:figure4}a, we note that there is a data point in the black trace that goes below zero.  This does not literally represent negative probability.  Rather, this is an artifact of how we calibrate the measurement signals that correspond to the ground state and the A mode's first excited state and how we use that calibration to convert averaged signals into probabilities.  When we perform an experiment, we program our room temperature equipment to repeatedly send a set of pulse sequences.  Some of these pulse sequences are used to calibrate our measured signals: we either prepare the qubits in the ground state (we do so passively by simply waiting the qubits to decay to their ground state) or the first excited state (by sending a $\pi$ pulse) and then send a measurement pulse.  We collect and average the resulting measurement signal for either the ground state or first excited state.  For each of the other pulse sequences (i.e. those corresponding to the actual experiment with pulse sequences associated with, say, Ramsey experiments or benchmarking experiments), we take the measured signal and project it onto an imaginary line in the I-Q plane that passes through the calibrated results for either the ground state or first excited state.  For a given pulse sequence, we take the average signal from multiple repetitions.  The projected position along that line is used to convert the averaged signal to a probability $P(|1\rangle)$ of being in the first excited state.  Being close to the calibrated ground state measurement means $P(|1\rangle)$ is close to zero and being close to the calibrated first excited state measurement means $P(|1\rangle)$ is close to 1; measured signals that are exactly halfway between the two calibrated points are assigned $P(|1\rangle) = 0.5$.   Crucially, because there is random/statistical scatter in our measurement points, sometimes a particular measured signal does not fall exactly between the calibrated points in I-Q space.  Our algorithm then assigns a probability value that naively seems to be non-physical, but this merely represents noise in our signal.

In Fig.~\ref{fig:figure4}b and c, we show the results of interleaved randomized benchmarking \cite{PhysRevLett.109.080505} for the echoed RIP gate between the A modes of Q0 and Q2.  We once again chose a bus drive detuning of $-20$ MHz and an amplitude corresponding to a total gate of 740 nanoseconds (which includes the time for the two RIP pulses, the echo pulses, and time buffers inserted before and after the RIP pulses).  We observe an error per clifford of 2.4\% for the reference sequences with an average of 1.5 RIP gates per Clifford (red dots).  Because we used a particular decomposition with an average of 1.5 RIP gates per Clifford, this translates to an upper bound of 1.7\% error per RIP gate.  While this is consistent with the error per gate of 1.3\% extracted from the interleaved randomized benchmarking (as determined by subtracting the EPC of the red reference data from the EPC of the blue data with an extra RIP gate per Clifford), it is inconsistent with the coherence limited error rate of 2.0\%.  Although the origin of this small discrepancy is uncertain, it could be due to temporal coherence fluctuations or the multiple single-qubit pulses during randomized benchmarking help to echo out low frequency noise that is not completely removed by the single echo pulse during a measurement of $T_{2,echo}$ that is used to calculate the coherence limited error.

Before closing, we expect that increasing the coupling capacitance between the bus resonator and the TCQs could increase the longitudinal coupling between the A modes and the bus resonator.  This will correspondingly increase the gate speed and allow for shorter RIP gates with higher fidelity.

\dograph{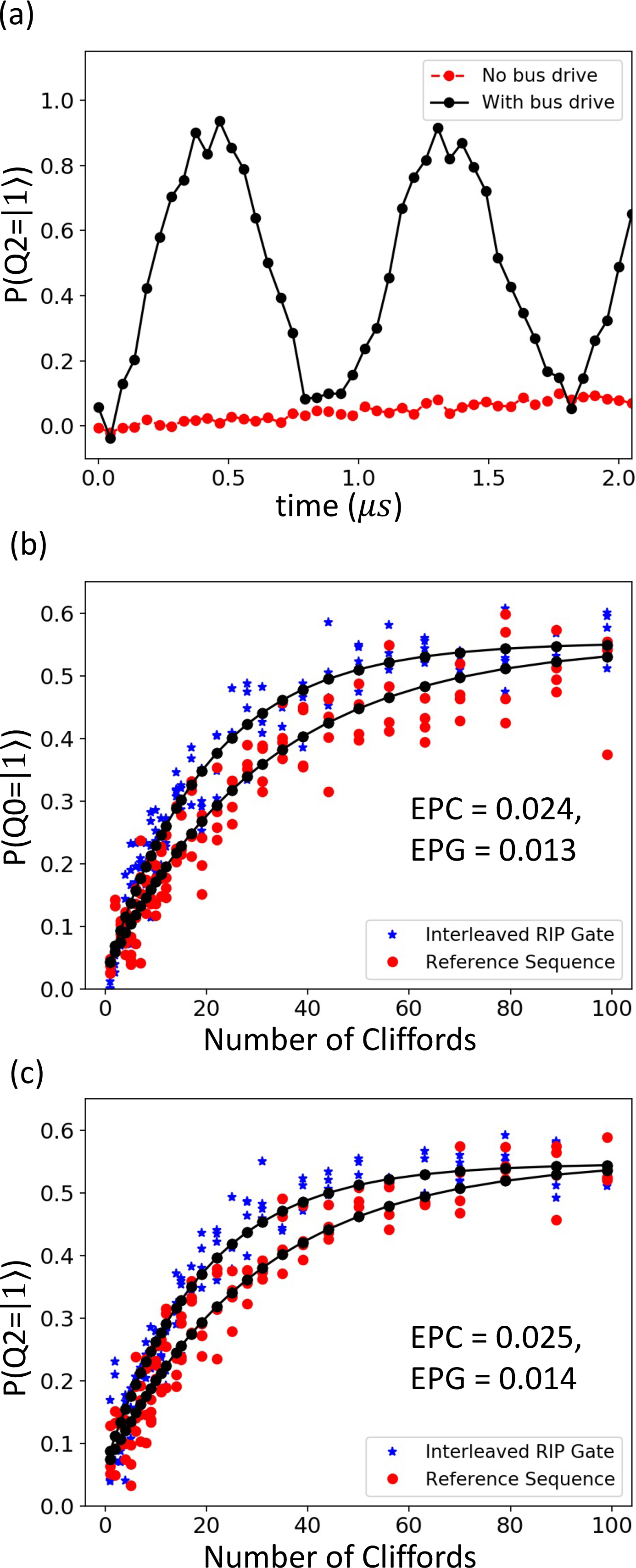}{width = 3.5 in}{(a) Ramsey experiment of Q2, with (black) and without (red) a drive applied to the bus between Q0 and Q2. (b) and (c) Interleaved randomized benchmarking of RIP gate between Q0 and Q2.  Blue stars are random sequences with interleaved RIP gates, red dots are reference sequences with an average of 1.5 RIP gates per Clifford, and black dots/lines are exponential fits.  We report the EPC (error per Clifford) and interleaved EPG (error per gate).}

\section{Conclusion}

In conclusion, we have demonstrated an architecture comprising of mode-selective coupling between two-junction qubits.  We have shown that there is dramatically suppressed static ZZ between the data modes for a wide range of detuning.  Despite the absence of exchange coupling between the bus resonators and the data modes, there is a sizable longitudinal coupling between them, which permits an all-microwave RIP gate between data modes.

$\it{Acknowledgements.}$ We thank D.~McKay, J.~Raftery and W.~Shanks for helpful discussions.  We acknowledge P.~Gumann for cryostat assistance.

\bibliography{tcq_rip}

\end{document}